\documentclass[11pt]{amsart}
\usepackage{amsmath,amstext,amsthm,amsfonts,amssymb,amsthm}
\usepackage{multicol}
\usepackage{latexsym}
\usepackage{color}
\usepackage{graphicx}
\usepackage{epsf}
\usepackage{epsfig}
\usepackage{subfigure}
\usepackage{rotating}
\usepackage{curves}
\usepackage{epic}

\usepackage{graphics}
\usepackage{verbatim}
\usepackage{color}
\usepackage{hyperref}

\newtheorem{theorem}{Theorem}[section]
\newtheorem{lemma}[theorem]{Lemma}
\theoremstyle{definition}

\theoremstyle{remark}
\newtheorem{remark}[theorem]{Remark}
\numberwithin{equation}{section}

\newcommand{\N}{\mathcal{N}}

\newcommand{\HI}{\mathfrak{H}}

\newcommand{\D}{\mathcal{D}}
\newcommand{\C}{\mathbb{C}}

\newcommand{\PP}{P_{m,n}^{\alpha}}
\newcommand{\pp}{\mathfrak{p}_{m,n}^{\alpha}}

\newcommand{\oz}{\overline{z}}

\begin{document}
\title[Quantization of the disc ]{2D-Zernike polynomials and coherent state quantization of the unit disc}
\author{K. Thirulogasanthar$^1$, Nasser Saad$^2$ and G. Honnouvo$^3$ }
\address{$^{1}$ Department of Computer Science and Software
Engineering, Concordia University, 1455 De Maisonneuve Blvd. West,
Montreal, Quebec, H3G 1M8, Canada. }
\address{$^2$ Department of mathematics and Statistics, University of Prince Edward Island, 550 University avenue, Charlottetown, UPEI,  C1A 4P3, Canada.}
\address{$^3$ Department of Mathematics and Statistics, McGill University, 805 Sherbrooke Street w., Montreal, Quebec, H3A 2K6, Canada}
\email{santhar@gmail.com, nsaad@upei.ca and g\_honnouvo@yahoo.fr}
\thanks{The research of one of the authors (NS)  was supported by the Natural Science and Engineering Research Council of Canada (NSERC)}
\date{\today}

\begin{abstract}
Using the orthonormality of the 2D-Zernike polynomials, reproducing kernels, reproducing kernel Hilbert spaces, and ensuring coherent states attained.  With the aid of the so-obtained coherent states, the complex unit disc is quantized. Associated upper symbols, lower symbols and related generalized Berezin transforms also obtained.  A number of necessary summation formulas for the 2D-Zernike polynomials proved.
\end{abstract}

\maketitle
\pagestyle{myheadings}

\section{Introduction}\label{sec_intro}
\noindent Quantization is commonly understood as a transition from classical to quantum formalisms. To a certain extent, quantization relates to a larger discipline than just restricting to a specific domain of applications.  In physics \cite{Gaz}, quantization is a procedure that associates with an algebra $A_{cl}$ of classical observables an algebra $A_q$ of quantum observables. The algebra $A_{cl}$ is usually realized as a commutative Poisson algebra of derivable functions on a symplectic (or phase) space $X$. The algebra $A_q$ is, however, noncommutative in general and the quantization procedure must provide a correspondence $A_{cl}\mapsto A_q:f\mapsto A_f$. Most quantum theories may be obtained as the result of a canonical quantization procedure. However, among the various quantization procedures available in the literature, the coherent state quantization (CS quantization) appear quite arbitrary because the only structure that a space $X$ must possess is a measure. Once a family of CS or frame, labelled by a measure space $X$, is given, one can quantize the measure space $X$. Various quantization schemes, advantages and drawbacks are discussed in detail in \cite{AE,Gaz, Per, Ali}.
\vskip0.1true in

\noindent In CS quantization, a correspondence between classical and quantum observables, usually, provided through a suitable generalization of the standard CS. The CS quantization is judged as equivalent to canonical quantization on a physical level. Recently however, a family of CS were obtained \cite{Nic} using the complex Hermite polynomials $H_{m,n}(z,\oz)$ wherein it was used to quantize the complex plane.  These results were further investigated and refined in \cite{ABG}. The results so obtained were departed from the results of the standard canonical quantization. Further, in \cite{Ga}, the CS, obtained using the complex Hermite polynomials $H_n(z)$, were used to quantize the so-called noncommutative plane. A similar approach was also used in \cite{AGH} to quantize the cylindrical phase spaces.
\vskip0.1true in

\noindent Following these recent developments, we briefly review, in Section 2, the general scheme of  CS quantization. Section 3 describes the physical relevance of the unit disc and the standard CS on the unit disc.  In Section 4, we present the 2D-Zernike polynomials, wherein we prove several summation formulas and an integral formula that are essential in obtaining reproducing kernels, CS, lower symbols and associated Berezin transforms. In Section 5, we derive a class of reproducing kernels from the 2D-Zernike polynomials and related reproducing kernel Hilbert spaces and also acquire their direct sum as an $L^2$ space.  Alongside, we also consider some projection operators and their integral Schwartz kernels. In Section 6,  we implement the CS quantization of the unit disc (also known as Lobachevsky plane) using the general scheme described in Section 2 and obtain the associated upper symbols, lower symbols and a generalized Berezin transform. The unit disc is the natural phase space for quantum systems with $su(1,1)$ symmetry \cite{Gaz}. However, we show that the operators arising from the 2D-Zerkine polynomial quantization described in this note do not obey the $su(1,1)$ symmetry (see Remark (\ref{rem1})). We end this manuscript with a conclusion in section 7.
\section{Coherent state quantization: General scheme}
\noindent Let $(X,\mu)$ be a measure space and
$$L^2(X,\mu)=\left\{f:X\longrightarrow\mathbb{C}~\vert~\int_X|f(x)|^2d\mu(x)<\infty\right\},$$
the Klauder-Berezin or anti-Wick or Toeplitz or coherent state quantization \cite{Nic,Hall}, as used by various authors, associates a classical observable that is a function $f(x)$ on $X$ to an operator valued integral. We continue with the general procedure described in \cite{Gaz} and used, for example, in the published work \cite{AGH, Nic,Ga}.
\vskip0.1true in
\noindent Choose a countable orthonormal family $\mathcal{O}=\{\phi_n~\vert~n\in\mathbb{N}\}$ in $L^2(X,\mu)$,
\begin{equation}\label{E1}
\langle\phi_n|\phi_m\rangle=\int_X\overline{\phi_m(x)}\,\phi_n(x)\,d\mu(x)=\delta_{mn}
\end{equation}
and assume that
\begin{equation}\label{E2}
0<\sum_{n=0}^{\infty}|\phi_n(x)|^2:=\mathcal{N}(x)<\infty\quad a.e.
\end{equation}
holds. Let $\mathfrak{H}$ be a separable complex Hilbert space with orthonormal basis $\{|e_n\rangle~\vert~n\in\mathbb{N}\}$ in 1-1 correspondence with $\mathcal{O}$. In particular, $\mathfrak{H}$ can be taken as $\mathcal{K}_0=\overline{\text{span}~\mathcal{O}}$ in $L^2(X,\mu)$. Then the family $\mathcal{F}_{\mathfrak{H}}=\{|x\rangle~\vert~x\in X\}$ with
\begin{equation}\label{E3}
|x\rangle=[\mathcal{N}(x)]^{-\frac{1}{2}}\sum_{n=0}^{\infty}\overline{\phi_n(x)}|e_n\rangle\in\mathfrak{H}
\end{equation}
forms a set of coherent states\,(CS). Among the numerous properties of these states \cite{Gaz}, we recall the following two features, namely normalization and resolution of identity relations
\begin{eqnarray}
& &\langle x\vert x\rangle_{\mathfrak{H}}=1,\label{E4}\\
& &\int_X\mathcal{N}(x)|x\rangle\langle x|d\mu(x)=\mathbb{I}_{\mathfrak{H}}.\label{E5}
\end{eqnarray}
 Equation (\ref{E5}) allows us to implement CS or frame quantization of the set of parameters $X$ by associating a function
$$X\ni x\mapsto f(x),$$
that satisfies appropriate conditions, the following operator in $\mathfrak{H}$
\begin{equation}\label{E6}
f(x)\mapsto A_f=\int_X\mathcal{N}(x)\,f(x)\,|x\rangle\langle x|\,d\mu(x).
\end{equation}
The matrix elements of $A_f$ with respect to the basis $\{|e_n\rangle\}$ are give by
\begin{equation}\label{E7}
\left(A_f\right)_{mn}=\langle e_m\vert A_f\vert e_n\rangle=\int_Xf(x)\,\overline{\phi_m(x)}\,\phi_n(x)\,d\mu(x).
\end{equation}
The operator $A_f$ is
\begin{enumerate}
\item[(a)]symmetric, if $f(x)$ is real valued.
\item[(b)]bounded, if $f(x)$ is bounded.
\item[(c)]self-adjoint, if $f(x)$ is real semi-bounded (through Friedrich's extension).
\end{enumerate}
In order to view the upper symbol $f$ of $A_f$ as a quantizable object (with respect to the family $\mathcal{F}_{\mathfrak{H}}$) a reasonable requirement is that the so-called lower symbol of $A_f$ defined as
\begin{equation}\label{E8}
\check{f}(x)=\langle x\vert A_f\vert x\rangle=\int_X\mathcal{N}(x')f(x')|\langle x\vert x'\rangle|^2d\mu(x')
\end{equation}
be a smooth function on $X$ with respect to some topology assigned to the set $X$. Associating to the classical observable $f(x)$ the mean value $\langle x|A_f|x\rangle$, one can also get the so-called Berezin transform $B[f]$ with
$B[f](x)=\langle x|A_f|x\rangle$, see for example \cite{Mo} for further details.
\section{Unit disc as a phase space}
\noindent We briefly review the necessary information associated with CS on the unit disc needed for the present work, for further detail we refer the reader to the standard monographs \cite{AE, Gaz,Per}. There exist many problems in theoretical physics wherein the unit disc, $\D=\{z\in\C~\vert~~|z|<1\}$, play an important role. For instance, it is a model of phase space for the motion of a material particle on a one sheeted two-dimensional hyperboloid viewed as a (1+1)-dimensional space-time with negative constant curvature, namely, the two dimensional {\em anti de Sitter} space-time. The unit disc equipped with a K\"ahlerian potential,
\begin{equation}\label{U1}
\mathcal{K}_{\D}(z,\oz)=\pi^{-1}(1-|z|^2)^2,
\end{equation}
has also the structure of a two-dimensional K\"ahlerian manifold. Any K\"ahlerian manifold is symplectic and so can be regard as a phase space for several mechanical systems.
\subsection{Standard CS on unit disc}
Let $\eta>1/2$ be a real parameter and let us equip the unit disc, $\D=\{z\in\C~\vert~~|z|<1\}$, with a positive measure
\begin{equation}\label{U2}
d\mu_{\eta}(z)=\frac{2\eta-1}{\pi}\frac{d^2z}{(1-|z|^2)^2}.
\end{equation}
Consider now the Hilbert space $L^2_{\eta}(\D,\mu_{\eta})$ of all functions $f(z,\oz)$ on $\D$ that are square integrable with respect to $\mu_{\eta}$. Let
\begin{equation}\label{U3}
B=\left\{\phi_n(z,\oz)=\sqrt{\frac{(2\eta)_n}{n!}}(1-|z|^2)^{\eta}~\oz^n~~\vert~~n\in\mathbb{N}\right\},
\end{equation}
where $(2\eta)_n$ is the Pochhammer symbol defined in terms of Gamma function as $(2\eta)_n=\Gamma(2\eta+n)/\Gamma(2\eta)$. Then, $B$ is an orthonormal system of $\mathcal{K}_+=\overline{\text{span}(B)}$ satisfying
\begin{equation}\label{U4}
\sum_{n=0}^{\infty}\left\vert\phi_n(z,\oz)\right\vert^2=1.
\end{equation}
Thereby we can readily write a set of CS
\begin{equation}\label{U5}
\vert z,\eta\rangle=\sum_{n=0}^{\infty}\overline{\phi_n(z,\oz)}|e_n\rangle
=(1-|z|^2)^{\eta}\sum_{n=0}^{\infty}\sqrt{\frac{(2\eta)_n}{n!}}z^n|e_n\rangle,
\end{equation}
where $\{|e_n\rangle|n\in\mathbb{N}\}$ is an orthonormal basis of a separable Hilbert space $\HI$. Note that, the coherent states \eqref{U5} are the same as those built by Perelomov for $SU(1,1)$ when the {\em fiducial vector} is chosen as $|p_0\rangle$ or $|e_0\rangle$. By construction
\begin{equation}\label{U6}
\langle z,\eta\vert z,\eta\rangle_{\HI}=1;\quad \int_Dd\mu_{\eta}(z)|z,\eta\rangle\langle z,\eta|=I_{\HI}.
\end{equation}
The Fock-Bargmann space $\mathcal{K}_+$ is also a reproducing kernel Hilbert space with reproducing kernel
\begin{equation}\label{U7}
K(z,w)=(1-z\oz)^{\eta}(1-w\overline{w})^{\eta}/(1-\oz w)^{2\eta}.
\end{equation}
For a given $\eta>1$, we introduce the Fock-Bargmann Hilbert space $\mathcal{FB}_{\eta}$ of all analytic functions $f_k(z)$ on $\D$ that are square integrable with respect to the scalar product
\begin{equation}\label{U8}
\langle f_1\vert f_2\rangle=\frac{2\eta-1}{2\pi}\int_{\D}\overline{f_1(z)}f_2(z)(1-|z|^2)^{2\eta-2}d^2z.
\end{equation}
Then
\begin{equation}\label{U9}
B_{\eta}=\left\{p_n(\xi)=\langle\xi|p_n\rangle=\sqrt{\frac{(2\eta)_n}{n!}}\xi^n~\vert~n\in\mathbb{N}\right\}
\end{equation}
is an orthonormal basis of $\mathcal{FB}_{\eta}$. 
Now, viewing the CS in (\ref{U5}) as vectors in $\mathcal{FB}_{\eta}$, we have
\begin{equation}\label{U10}
\vert z,\eta\rangle=(1-|z|^2)^{\eta}\sum_{j=0}^{\infty}\sqrt{\frac{(2\eta)_j}{j!}}z^j|p_j\rangle.
\end{equation}
\section{2D-Zernike polynomials on the unit disc}
\noindent The family of 2D-Zernike polynomials is well studied, for example see \cite{Wu3}, and some interesting applications, in quantum optics, recently investigated in \cite{Tor}. Follows the notations of A. W\"unsche \cite{Wu3}, let $z=re^{i\theta}, \oz=re^{-i\theta}$, and  $\D=\{z\in\C~\vert~ |z|^2=z\oz<1\}$ be the complex unit disc. The 2D-Zernike polynomials are defined by \cite{Wu3}
\begin{align}\label{hy1}
\PP(z,\oz)&=\dfrac{n!}{(\alpha+1)_n}z^{m-n} {\mathcal P}_n^{(\alpha,m-n)}(2z\oz-1)=\dfrac{m!}{(\alpha+1)_m}\oz^{n-m} {\mathcal P}_m^{(\alpha,n-m)}(2z\oz-1)\notag\\
&=z^m\,\oz^n\, {}_2F_1(-m,-n;\alpha+1;1-\frac{1}{z\oz});\quad\alpha>-1, m,n=0,1,2,...,
\end{align}
where ${}_2F_1(a,b;c;z)$ is the hypergeometric function and ${\mathcal P}_l^{(n,m)}(z) $ represents the Jacobi polynomials. Clearly, $\PP(z,\oz)$ are polynomials of $(z,\oz)$ of $(m+n)$-degree satisfying the following properties,
\begin{equation}\label{E9}
\PP(-z,-\oz)=(-1)^{m+n}\PP(z,\oz);\quad\overline{\PP(z,\oz)}=\PP(\oz,z)=P_{n,m}^{\alpha}(z,\oz).
\end{equation}
Using the area element of the plane as
$$\frac{i}{2}dz\wedge d\oz=rdr\wedge d \phi=dx\wedge dy,$$
we have the orthogonality relation \cite{Wu3}
\begin{equation}\label{E10}
\int_{z\oz\leq 1}\frac{i}{2}\,dz\wedge d\oz\,(1-z\oz)^{\alpha}\,\overline{P_{k,l}^{\alpha}\,(z,\oz)}\PP(z,\oz)=A_{\alpha}(m,n)\delta_{km}\delta_{ln},
\end{equation}
where
\begin{equation}\label{ES10}
A_{\alpha}(m,n)=\frac{\pi\, m!\,n!}{(m+n+\alpha+1)\,(\alpha+1)_m\,(\alpha+1)_n},\quad \alpha>-1.
\end{equation}
Introduce 
\begin{equation}\label{E11}
\pp(z,\oz)=[A_{\alpha}(m,n)]^{-\frac{1}{2}}(1-z\oz)^{\frac{\alpha}{2}}\PP(z,\oz),
\end{equation}
then
\begin{equation}\label{E12}
\langle \mathfrak{p}^{\alpha}_{k,l}\vert\pp\rangle=\int_{z\oz\leq 1}\frac{i}{2}dz\wedge d\oz~\overline{\mathfrak{p}^{\alpha}_{k,l}(z,\oz)}\pp(z,\oz)=\delta_{km}\delta_{ln}
\end{equation}
defines an inner product and the set $B_{\alpha}=\{\pp(z,\oz)~\vert~m,n=0,1,2...;\alpha~\text{fixed}\}$ is complete in $L^2_{\alpha}(\D,\frac{i}{2}dz\wedge d\oz)$. In which case, the completeness relation \cite{Wu3} reads
\begin{equation}\label{E13}
\sum_{m=0}^{\infty}\sum_{n=0}^{\infty}\pp(z,\oz)\,\overline{\pp(z',\oz')}=\delta(z-z',\oz-\oz');\quad |z|,|z'|<1,
\end{equation}
where $\delta(x,y)$ is the two-dimensional Dirac delta function.
Among the recurrence relations of $\PP(z,\oz)$ provided in \cite{Wu3}, we also mention
\begin{equation}\label{RR1}
(m+n+\alpha+1)z\PP(z,\oz)=(m+1+\alpha)P_{m+1,n}^{\alpha}(z,\oz)+nP_{m,n-1}^{\alpha}(z,\oz).
\end{equation}
Our main results are summarized in the following lemmas that are essential in obtaining a reproducing kernel Hilbert space and, thereby, CS and to the computations of lower symbols and in so establishing Berezin transforms. To the best of our knowledge, the following summation formulas and the integral formula for the 2D-Zernike polynomials have not yet been worked out. 
\begin{lemma}\label{T1}
For $\alpha>-1,~z\,\overline{z}<1$ and $w\,\overline{w}<1$, 
\begin{align}\label{S1}
\sum_{m=0}^{\infty}&\frac{\PP(z,\oz)\overline{\PP(w,\overline{w})}}{A_{\alpha}(m,n)}\notag\\
&= \dfrac{n(\alpha+1)_n(w-z)^n(\oz-\overline{w})^n}{\pi\,n!\,(1-z\overline{w})^{2n+\alpha+1}} {}_2F_1\left(-n,-n;1+\alpha;\dfrac{(z\oz-1)(w\overline{w}-1)}{(\oz-\overline{w})(w-z)}\right)\notag\\
&+\dfrac{(\alpha+1)_n\left(\dfrac{w}{z}\right)^n}{\pi\,n!\,(1-z\overline{w})^{\alpha+2}}\bigg[(\alpha+n+1)F_2\left(\alpha+2;-n,-n;\alpha+2,\alpha+1;\dfrac{z\oz-1}{z\overline{w}-1},\dfrac{z(w\overline{w}-1)}{w(z\overline{w}-1)}\right)\notag\\
&-nz\oz F_2\left(\alpha+2;1-n,-n;\alpha+2,\alpha+1;\dfrac{z\oz-1}{z\overline{w}-1},\dfrac{z(w\overline{w}-1)}{w(z\overline{w}-1)}\right)
\bigg]=E_{n}^{\alpha}(z,\overline{w})\quad\text{(say),}\notag\\
\end{align}
where $F_2$ is the second Appell function of two variables \cite{sri2}.
Particularly, we have
\begin{enumerate}
\item[(a)]$\displaystyle{\sum_{m=0}^{\infty}\frac{P_{m,0}^{\alpha}(z,\oz)\overline{P_{m,0}^{\alpha}(w,\overline{w})}}
{A_{\alpha}(m,0)}=\frac{(\alpha+1)}
{\pi(1-\overline{w}z)^{\alpha+2}}}$,
\item[(b)]$\displaystyle{\sum_{m=0}^{\infty}\frac{|\PP(z,\oz)|^2}{A_{\alpha}(m,n)}=\frac{(2n+\alpha+1)}
{\pi(1-\overline{z}z)^{\alpha+2}}};\quad n=0,1,2\cdots.$
\end{enumerate}
\end{lemma}
\begin{proof}
See the appendix.
\end{proof}
\noindent Similarly, it is straightforward to prove that
\begin{align}\label{S2}
\sum_{n=0}^{\infty}&\frac{\PP(z,\oz)\overline{\PP(w,\overline{w})}}{A_{\alpha}(m,n)}\notag\\
&=
\dfrac{m(\alpha+1)_m}{\pi\, m!}\dfrac{\left(z-w\right)^{m} \left(\overline{w}-\oz\right)^{m}}{
(1-w \overline{z})^{2m+1+\alpha}} {}_2F_1\left(-m,-m;1+\alpha;\dfrac{\left(z\oz -1\right) \left(w\overline{w}-1\right)}{\left(z-w\right) \left(\overline{w}-\oz\right)}\right)\notag\\
&+\dfrac{(\alpha+1)_m}{\pi\,m!(1-w\overline{z})^{\alpha+2}}\, \left(\dfrac{\overline{w}}{\oz}\right)^m\notag\\
&\times \bigg((1+\alpha+m)F_2(2+\alpha;-m,-m,a+2,a+1;\dfrac{z\oz-1}{\oz w-1},\dfrac{\oz(w\overline{w}-1)}{\overline{w}(w\overline{z}-1)})\notag\\
&-mz\oz F_2(2+\alpha;1-m,-m,a+2,a+1;\dfrac{z\oz-1}{\oz w-1},\dfrac{\oz(w\overline{w}-1)}{\overline{w}(w\overline{z}-1)})\bigg)
\equiv E_{m}^{\alpha}(\oz,w),\notag\\
\end{align}
particularly,
$$\displaystyle{\sum_{n=0}^{\infty}\frac{P_{0,n}^{\alpha}(z,\oz)\overline{P_{0,n}^{\alpha}(w,\overline{w})}}
{A_{\alpha}(0,n)}=\frac{(\alpha+1)}
{\pi(1-w\oz)^{\alpha+2}}};\quad \displaystyle{\sum_{n=0}^{\infty}\frac{|\PP(z,\oz)|^2}{A_{\alpha}(m,n)}=\frac{(2m+\alpha+1)}
{\pi(1-\overline{z}z)^{\alpha+2}}}.$$

\begin{lemma}\label{T2}
For each $n\in\mathbb{N}$ and $\alpha>-1,~z\oz<1$, $\eta>(\alpha+2)/2$, the series
\begin{equation}
\sum_{m=0}^\infty \dfrac{P_{m,n}^\alpha (z,\overline{z})}{\sqrt{A_\alpha(m,n)}}\sqrt{\dfrac{(2\eta)_m}{m!}}\xi^m,\quad n=0,1,2,\dots,
\end{equation}
is absolutely convergent. For the special case $n=0$ and $\eta=(2+\alpha)/2$, 
\begin{equation}
\sum_{m=0}^\infty \dfrac{P_{m,0}^\alpha (z,\overline{z})}{\sqrt{A_\alpha(m,0)}}\sqrt{\dfrac{(2+\alpha)_m}{m!}}\xi^m
=\sqrt{\frac{\alpha+1}{\pi}}(1-\xi z)^{-\alpha-2}.
\end{equation}
\end{lemma}
\begin{proof}
See the appendix.
\end{proof}
\begin{lemma}\label{T3}
Let $z=re^{i\theta}$. For $\alpha>-1$, $z\oz<1$ and $\theta\in[0,2\pi)$, we have
\begin{align*}
\int_{z\oz\leq 1}f(\theta) \,\overline{P_{l,n}^{\alpha}\,(z,\oz)}\PP(z,\oz)d\mu_\alpha(z,\oz)= \dfrac{C m!\,n!}{2(m+n+\alpha+1)(1+\alpha)_m(1+\alpha)_n}\delta_{l,m},
\end{align*}
where $C=\int_0^{2\pi}f(\theta)d\theta.$
\end{lemma}
\begin{proof}
See the appendix.
\end{proof}
\begin{lemma}\label{LL1}
For $\alpha>-1$, $z\oz<1$ we have
\begin{align*}
\int_{\mathcal{D}}&|z|^2 P_{m,n}^{\alpha}(z, \oz)\overline{P_{m,n}^{\alpha}(z, \oz)}d\mu_{\alpha}(z,\oz)
=\dfrac{\pi (m(m+1)+n(n+1)+\alpha(n+m+1))\,m!\,n!}{(\alpha+m+n)_3(\alpha+1)_m(\alpha+1)_n}.
\end{align*}
\end{lemma}
\begin{proof}
See the appendix.
\end{proof}
\begin{remark}
In Equation (\ref{S1}), if we set $n=0$ and $\alpha=0$, we get the usual Bergman kernel for the open unit disc $\D$, see for example \cite{Ko},
$$B(z,w)=\frac{1}{\pi(1-z\overline{w})^2};\quad z,w\in\D$$
and if we only set $n=0$ we get the weighted Bergman kernel \cite{E}
$$B_{\alpha}(z,w)=\frac{\alpha+1}{\pi(1-z\overline{w})^{\alpha+2}};\quad z,w\in\D$$
of the weighted Bergman space $\mathcal{A}_{\alpha}^2(\D)$.
\end{remark}
\section{Reproducing kernel Hilbert spaces}
\noindent In this section, we study the reproducing kernel Hilbert space associated with the 2D-Zernike polynomials $P_{m,n}^{\alpha}(z,\oz)$ for fixed $m$ and for fixed $n$ with $\alpha>-1$.  
We shall also consider the Schwartz kernel of some projection operators. In particular, we focus on the cases $n=0$ and $m=0$.
\vskip0.1true in
\noindent Let $(X, \nu)$ be a measure space. Whenever we have an orthonormal family $\left\{\Phi_m(x)\right\}_{m=0}^{\infty}$ with respect to the measure $d\nu$, satisfying 
\begin{equation}\label{RX1}
\displaystyle\sum_{n=0}^{\infty}|\Phi_n(x)|^2<\infty;\quad x\in X,
\end{equation}
we can readily form a reproducing kernel $$K(x,y)=\sum_{n=0}^{\infty}\overline{\Phi_n(x)}\Phi_n(y)$$ with the reproducing kernel Hilbert space $\mathfrak{H}_K=\overline{\text{span}}\{\Phi_n(x)~|~n=0,1,2,\cdots\}$, where the bar stands for the closure \cite{Ali}. By construction, it can be easily seen that the reproducing kernel $K(x,y)$ satifies the following properties.
\begin{enumerate}
\item[(a)]Hermiticity, $\displaystyle K(x,y)=\overline{K(y,x)}$ for all $x,y\in X$.
\item[(b)]Positivity, $\displaystyle K(x,x)\geq 0$ for all $x\in X$.
\item[(c)]Idempotence, $\displaystyle\int_XK(x,y)K(y,z)d\nu(y)=K(x,y)$ for all $x,y,z\in X$.
\end{enumerate}
In view of equations (\ref{E1}) and (\ref{E2}), the orthonormality of the family $\left\{\Phi_m(x)\right\}_{m=0}^{\infty}$ and the condition (\ref{RX1}) are the essential ingredients to obtain CS and thereby quantization. Further, if we take the CS as vectors in the space $\mathfrak{H}_K$ then the quantized operators will be operators from $\mathfrak{H}_K$ to $\mathfrak{H}_K$ (as if we replace $|e_n\rangle$ in (\ref{E3}) by $\phi_n$ then the vectors in (\ref{E3}) will be vectors in $\mathcal{K}_0$). Also note that the normalization factor $\mathcal{N}(x)$ of the CS in (\ref{E3}) is in fact the kernal $K(x,x)$. 
\vskip0.1true in

\noindent In this regard, we consider the following family 
$$B_m^{\alpha,n}(z,\oz)=\frac{\overline{P_{m,n}^{\alpha}(z,\oz)}}{\sqrt{A_{\alpha}(m,n)}}$$
where $P_{m,n}^{\alpha}(z,\oz)$ is the  2D-Zernike polynomials \eqref{hy1} and $A_\alpha(m,n)$ defined by \eqref{ES10}. For each $\alpha>-1$, let
$$d\mu_{\alpha}(z,\oz)=\frac{i}{2}(1-z\oz)^{\alpha}dz\wedge d\oz.$$
\begin{enumerate}
\item[$\bullet$]
According to (4.5), the family $B_{\alpha}=\left\{\overline{B_m^{\alpha,n}(z,\oz)}~|~m,n\in\mathbb{N}\right\}$ is orthonormal with respect to the measure $d\mu_{\alpha}$. However, because of \eqref{E13}, the necessary condition of existence \eqref{RX1} is not satisfied by the orthonormal family $B_{\alpha}$. Therefore, it cannot be possible to construct a reproducing kernel in this case.
\item[$\bullet$] For fixed $m$, according to \eqref{E10}, the family $B_m^{\alpha}=\left\{\overline{B_m^{\alpha,n}(z,\oz)}~|~{n\in\mathbb{N}}\right\}$ is orthonormal with respect to the measure $d\mu_{\alpha}$. Therefore, from lemma \ref{T2}, the corresponding reproducing kernel reads
 $$K_{\alpha}^m(z,w)=\sum_{n=0}^{\infty}{B_m^{\alpha,n}}(z,\oz)\overline{B_m^{\alpha,n}(w,\overline{w})}
=E_{m}^{\alpha}(\oz,w)$$
and the reproducing kernel Hilbert is $$\mathfrak{H}_K^{\alpha,m}=\overline{\text{span}}\left\{\overline{B_m^{\alpha,n}}~|~n=0,1,\cdots\right\}.$$ The case $m=0$, that is the space $\mathfrak{H}_K^{\alpha,0}=\overline{\text{span}}\left\{{z^n}/{\sqrt{A_{\alpha}(0,n)}}~|~n=0,1,\cdots\right\}$, corresponds to the holomorphic sector of  $L_{\alpha}^2(\mathcal{D}, d\mu_{\alpha}(z,\oz))$ with the reproducing kernel (see lemma \ref{T2})
$$K_{\alpha}^0(z,w)=\sum_{n=0}^{\infty}{B_0^{\alpha,n}}(z,\oz)\overline{B_0^{\alpha,n}(w,\overline{w})}=
\frac{(\alpha+1)}
{\pi}(1-w\oz)^{-\alpha-2}.$$
\item[$\bullet$] For fixed $n$, let
\begin{equation}\label{B1}
B_n^{\alpha}=\left\{\frac{\overline{\PP(z,\oz)}}{\sqrt{A_{\alpha}(m,n)}}~\vert~m=0,1,2,...\right\}.
\end{equation}
According to Lemma \ref{T1} the condition (\ref{RX1}) is satisfied by the orthonormal family $B_n^{\alpha}$. Let $\Lambda_n^{\alpha}(\D)=\overline{\text{span}~B_n^{\alpha}}$. Then $B_n^{\alpha}$ is a basis of $\Lambda_n^{\alpha}(\D)$ and it is a reproducing kernel Hilbert space with the kernel $E_n^{\alpha}(z,\overline{w})$ (see lemma \ref{T1}), we also need the following in the sequel 
\begin{equation}\label{R1}
\N_n(z,\oz)=E_n^{\alpha}(z,\oz)=\frac{(2n+\alpha+1)}{\pi}(1-z\overline{z})^{-\alpha-2};\quad n=0,1,2,\cdots. 
\end{equation}
Further, one can write
\begin{equation}\label{E23}
\displaystyle L_{\alpha}^2(\D, d\mu(z,\oz))=\bigoplus_{n=0}^{\infty}\Lambda_n^{\alpha}(\D).
\end{equation}
Thus 
$$B_0^{\alpha}=\left\{\frac{\oz^m}{\sqrt{A_{\alpha}(m,0)}}~\vert~m=0,1,2,...\right\},$$
is the basis of the space $\Lambda^{\alpha}_0(\D)$ and it is the classical Bargmann space of anti-holomorphic functions on the unit disc with reproducing kernel  
$$E_0^{\alpha}(z,\overline{w})=\frac{\alpha+1}{\pi}(1-\overline{w}z)^{-\alpha-2}.$$
Now for a given integral linear operator of the form
$$(Af)(x)=\int K_A(x,y)f(y)dy$$
the function $K_A(x,y)$ is called its Schwartz kernel \cite{ES}. As it was done in \cite{In}, here by constructions $B_n^{\alpha}$ is an orthonormal basis for $\Lambda_n^{\alpha}(\D)$ and are pairwise orthogonal in the Hilbert space $L_{\alpha}^2(\D, d\mu_{\alpha}(z,\oz))$, for the projection operator
$$P_n^{\alpha}:L^2_{\alpha}(\D, d\mu_{\alpha}(z,\oz))\longrightarrow \Lambda_n^{\alpha}(\D)$$
the integral Schwartz kernel $K_n^{\alpha}(z,w)$ can be obtained as
\begin{eqnarray*}
\left[P_n^{\alpha}f\right](z)&=&\sum_{m=0}^{\infty}\left\langle f\bigg|\frac{\PP}{\sqrt{A_{\alpha}(m,n)}}\right\rangle\frac{\PP(z,\oz)}{\sqrt{A_{\alpha}(m,n)}}\\
&=&\int_{\D}f(w)\sum_{m=0}^{\infty}\frac{\PP(z,\oz)\overline{\PP(w,\overline{w})}}{A_{\alpha}(m,n)}d\mu_{\alpha}(w,\overline{w}).
\end{eqnarray*}
Thereby the integral Schwartz kernel of the projection operator, $P_n^{\alpha}$, is given by
\begin{equation}\label{SW1}
K_n^{\alpha}(z,w)=\sum_{m=0}^{\infty}\frac{\overline{\PP(w,\overline{w})}\PP(z,\oz)}{A_{\alpha}(m,n)}
=E_n^{\alpha}(z,\overline{w}).\end{equation}
\end{enumerate}
\begin{remark}
It is worth mentioning that, in \cite{As, El}, the spaces $A_{\beta, n}(\mathcal{D})$ and 
$A_n^{2,m}(\mathcal D) ;~~n=0,1,2...$ are realized as the eigenspaces of the hyperbolic Landau levels with orthogonal basis expressed in terms of the Gauss hypergeometric functions, $_2F_1$. In particular, in \cite{As, El},  the authors have studied the CS and photon counting probabilities associated with the spaces  $A_{\beta,n}(\mathcal{D})$. For 
the space $A_{\beta, n}(\mathcal{D})$, the orthogonal basis is the set of functions (see \cite{As}, equations (3.7) and (3.8)),
\begin{align}\label{b1}
\Phi_{n}^{\beta, m}(z)
&=|z|^{|m-n|}e^{-i(m-n)\arg(z)}(1-|z|^2)^{-m}\notag\\
&\times {}_2F_1(-m+\dfrac{|m-n|-m+n}{2},2\beta-m+\dfrac{|m-n|-m+n}{2};1+|m-n|; |z|^2)\notag\\
\end{align}
while for the space $A_n^{2,m}(\mathcal D)$, the family of functions given explicitly in terms of the Jacobi polynomials as (see \cite{El}, equations (2.2) and (2.3))
\begin{align}\label{b2}
\Phi_{n}^{\beta, m}(z)
&=(-1)^{\min(m,n)}|z|^{|m-n|}e^{-i(m-n)\arg(z)}(1-|z|^2)^{-m} P_{\min(m,n)}^{(|m-n|,2(\beta-m)-1)}(1-2|z|^2)\notag\\
\end{align}
Both of these equivalent spaces are similar to the space $\Lambda_n^{\alpha}(\mathcal{D})$, where $$B_n^{\alpha}=(1-|z|^2)^m \Phi_{n}^{\beta, m}(z),\quad \alpha=2\beta-n-m-1, for~ fixed~m,  $$
up to a constant. Indeed, by means of Pfaff transformation of hypergeometric functions, it is not difficult to show, for $\alpha=2\beta-n-m-1$ and $m<n$, that
\begin{align*}
P_{m,n}^{2\beta-n-m-1} (z,\overline{z})&
=\overline{z}^{n-m}\dfrac{(-n)_m}{(2\beta-n-m)_m}{}_2F_1(-m,2\beta-m;1+n-m; z\overline{z})\\
&=|z|^{n-m}e^{-i(n-m)\arg(z)}\dfrac{(-n)_m}{(2\beta-n-m)_m}{}_2F_1(-m,2\beta-m;1+n-m; |z|^2)\\
\end{align*}
while for $n<m$
\begin{align*}
P_{m,n}^{2\beta-n-m-1} (z,\overline{z})&
=\overline{z}^{m-n}\dfrac{(-m)_n}{(2\beta-n-m)_n}{}_2F_1(-n,2\beta-n;1+m-n; z\overline{z})\\
&=|z|^{m-n}e^{-i(m-n)\arg(z)}\dfrac{(-m)_n}{(2\beta-n-m)_n}{}_2F_1(-n,2\beta-n;1+m-n; |z|^2)\\
\end{align*}
These  last two expressions can be written as
\begin{align*}
P_{m,n}^{2\beta-n-m-1} (z,\overline{z})
&=|z|^{|m-n|}e^{-i(m-n)\arg(z)} \dfrac{(-m)_n}{(2\beta-n-m)_n}\notag\\
&\times {}_2F_1(-m+\dfrac{|m-n|-m+n}{2},2\beta-m+\dfrac{|m-n|-m+n}{2};1+|m-n|; |z|^2).
\end{align*}
 For further details we refer the reader to \cite{As, El}. 
\end{remark}
\section{Quantization of the complex unit disc}
\noindent In order to adapt to the general construction, let us assume, for the disc $\D$, the follows:
\begin{enumerate}
\item[(a)]Let $\displaystyle d_{\alpha}\mu(z,\oz)=\frac{i}{2}(1-z\oz)^{\alpha}dz\wedge d\oz$.
\item[(b)]$\displaystyle (\D, d\mu_{\alpha}(z,\oz))$ is a measure space.
\item[(c)]$\displaystyle L^2_{\alpha}(\D, d\mu_{\alpha}(z,\oz))$ is the corresponding Hilbert space of complex valued square integrable functions.
\item[(d)]$\displaystyle\left\{{\overline{\PP(z,\oz)}}/{\sqrt{A_{\alpha}(m,n)}}~\vert ~ m=0,1,2,...;\alpha, n~~\text{fixed}\right\}$ is an orthonormal set in \\ $L^2_{\alpha}(\D, d\mu_{\alpha}(z,\oz))$.
\item[(e)]$\displaystyle\{|e_m\rangle~|~m=0,1,2,...\}$ be an orthonormal basis of an abstract Hilbert space $\HI$.
\end{enumerate}
\vskip0.1true in
With the above assumptions, consider the vectors
\begin{equation}\label{E20}
\vert z,\alpha,n\rangle=\N_n(z,\oz)^{-\frac{1}{2}}\sum_{m=0}^{\infty}
\frac{\PP(z,\oz)}{\sqrt{A_{\alpha}(m,n)}}|e_m\rangle.
\end{equation}
From (\ref{E12}) and lemma (\ref{T1})-(b), the vectors in (\ref{E20}) forms a set of CS in the sense
\begin{enumerate}
\item[$\bullet$]they are normalized, $\langle z,\alpha, n\vert z,\alpha, n\rangle=1$, and
\item[$\bullet$]satisfy a resolution of the identity
\begin{equation}\label{E21}
\int_{\D}\N_n(z,\oz)\vert z,\alpha,n\rangle\langle z,\alpha,n\vert d\mu_{\alpha}(z,\oz)=I_{\HI}.
\end{equation}
\end{enumerate}
Equation (\ref{E21}) allows us to implement CS quantization of the disc $\D$ by associating a function
$$\D\ni z\mapsto f(z,\oz).$$
For this define the operator on $\HI$
\begin{equation}\label{E22}
f(z,\oz)\mapsto A_f=\int_{\D}\N_n(z,\oz)f(z,\oz)\vert z,\alpha,n\rangle\langle z,\alpha,n\vert d\mu_{\alpha}(z,\oz).
\end{equation}
That is
\begin{equation}\label{E233}
A_f=\sum_{m=0}^{\infty}\sum_{l=0}^{\infty}\frac{|e_m\rangle\langle e_l|}{\sqrt{A_{\alpha}(m,n)A_{\alpha}(l,n)}}\int_{\D}f(z,\oz)\PP(z,\oz)\overline{P_{l,n}^{\alpha}(z,\oz)}d\mu_{\alpha}(z,\oz).
\end{equation}
From the orthogonality relation (\ref{E10}), for $f(z)=1$, we immediately get $A_1=I_{\HI}$. Further, using the recursion relation (\ref{RR1}) and the orthogonality relation (\ref{E10}), it can easily be seen that
\begin{eqnarray}\label{E24}
A_z&=&\sum_{m=0}^{\infty}C(m,n,\alpha)|e_m\rangle\langle e_{m+1}|\\
A_{\oz}&=&\sum_{m=0}^{\infty}C(m,n,\alpha)|e_{m+1}\rangle\langle e_{m}|
\end{eqnarray}
where
$$C(m,n,\alpha)=\sqrt{\frac{(m+1)(m+\alpha+1)}{(m+n+\alpha+2)(m+n+\alpha+1)}}.$$
Thereby, we have
\begin{eqnarray*}
A_z|e_0\rangle&=&0,\\
A_z|e_j\rangle&=&C(j-1,n,\alpha)|e_{j-1}\rangle; ~j=1,2,\cdots,\\
A_{\oz}|e_j\rangle&=&C(j,n,\alpha)|e_{j+1}\rangle; ~j=0,1,2,\cdots.
\end{eqnarray*}
That is $A_z, A_{\oz}$ are annihilation and creation operators, respectively. Their commutators takes the form
\begin{equation}\label{E25}
[A_z,A_{\oz}]=C^2(0,n,\alpha)|e_0\rangle\langle e_0|+\sum_{m=1}^{\infty}\left[C^2(m,n,\alpha)-C^2(m-1,n,\alpha)\right]|e_m\rangle\langle e_m|.
\end{equation}
It may be interesting to note that for $n=0$ and $\alpha=0$ it becomes
$$[A_z,A_{\oz}]=\sum_{m=0}^{\infty}\frac{|e_m\rangle\langle e_m|}{(m+1)(m+2)}.$$
In the basis $\{|e_m\rangle~|~m=0,1,2,\cdots\}$, the matrix elements read
$$(A_z)_{k,l}=\langle e_k|A_z|e_l\rangle=\left\{\begin{array}{ccc}0&\text{if}&l\not=k+1\\
C(k+1,n,\alpha)&\text{if}&l=k+1\end{array}\right.$$
and
$$(A_{\oz})_{k,l}=\langle e_k|A_{\oz}|e_l\rangle=\left\{\begin{array}{ccc}0&\text{if}&k\not=l+1\\
C(k-1,n,\alpha)&\text{if}&k=l+1\end{array}\right.$$
Since we are in the unit disc, another interesting pair is the quatization of $|z|^2$ and $\text{Arg}(z)=\theta$; $0\leq\theta<2\pi$. 
Using lemma \ref{LL1} we obtain
\begin{align*}
A_{|z|^2}&=\sum_{m=0}^{\infty}\frac{|e_m\rangle\langle e_m|}{A_{\alpha}(m,n)}\int_{\mathcal{D}}|z|^2 P_{m,n}^{\alpha}(z, \oz)\overline{P_{l,n}^{\alpha}(z, \oz)}d\mu_{\alpha}(z,\oz)\\
&=\sum_{m=0}^{\infty}|e_m\rangle\langle e_m|\frac{(m+n+\alpha+1)\,(\alpha+1)_m\,(\alpha+1)_n}
{\pi\,m!\,n!} \dfrac{\pi (m(m+1)+n(n+1)+\alpha(n+m+1))\,m!\,n!}{(\alpha+m+n)_3(\alpha+1)_m(\alpha+1)_n}\\
&=\sum_{m=0}^{\infty}|e_m\rangle\langle e_m|
\dfrac{(m(m+1)+n(n+1)+\alpha(n+m+1))}{(\alpha+m+n)(\alpha+m+n+2)}\\
&=\sum_{m=0}^{\infty}|e_m\rangle\langle e_m|\left(\frac{(m+1)(n+1)}{m+n+\alpha+2}-\dfrac{mn}{m+n+\alpha}.
\right)\end{align*}
For the quantization of $\theta$, using the integral formula in lemma \ref{T3}, we have
\begin{eqnarray*}
A_{\theta}&=&\sum_{m=0}^{\infty}\sum_{l=0}^{\infty}\frac{|e_m\rangle\langle e_l|}{\sqrt{A_{\alpha}(m,n)A_{\alpha}(l,n)}}\int_{\mathcal{D}}\theta P_{m,n}^{\alpha}(z, \oz)\overline{P_{l,n}^{\alpha}(z, \oz)}d\mu_{\alpha}(z,\oz)\\
&=&\sum_{m=0}^\infty \sum_{l=0}^\infty\dfrac{|e_m\rangle\langle e_l|}{\sqrt{A_\alpha(m,n)A_\alpha(l,n)}}\dfrac{\pi^2 m!\,n!}{(m+n+\alpha+1)(1+\alpha)_m(1+\alpha)_n}\delta_{l,m}.\notag\\
&=&\sum_{m=0}^\infty \dfrac{|e_m\rangle\langle e_m|}{A_\alpha(m,n)}\dfrac{\pi^2 m!\,n!}{(m+n+\alpha+1)(1+\alpha)_m(1+\alpha)_n}\\
&=&\pi \sum_{m=0}^\infty |e_m\rangle\langle e_m|
=\pi I_{\mathfrak{H}},
\end{eqnarray*}
\begin{remark}
An interesting feature appears from lemma 4.3. For any function $f(\theta)$, similar to  the calculation of $A_{\theta}$, it follows that (we place the proof of this assertion in the appendix:
\begin{eqnarray*}
A_{f(\theta)}&=&\frac{C}{2\pi} I_{\mathfrak{H}}, \qquad \mbox{where $C=\int_0^{2\pi}f(\theta) d\theta$},
\end{eqnarray*}
This means that the quantization map $A_f$ of equation (\ref{E233}) is insensitive to the variable $\theta$. It  is only sensitive to the average of functions of $\theta$.
\end{remark} 
\begin{remark}\label{rem1}
In \cite{Nic}, the authors have quantized the complex plane using the CS built from the $2D$- complex Hermite polynomials $H_{n+s,s}(z,\oz)$. In their construction $z\in\mathbb{C}$ and the operators $A_z$ and $A_{\oz}$ depend on the parameter $s$. By setting $s=0$ in the commutation relation of $A_z$ and $A_{\oz}$ the authors have recovered the usual canonical commutation relation, namely $[A_z, A_{\oz}]=I_{\mathfrak{H}}$. In the same spirit, since we have quantized the unit disc, a natural phase space for quantum systems with $su(1,1)$ symmetry, one may expect to recover the $su(1,1)$ commutation relation for some particular values of $\alpha$ and $n$ (note that in our case, $z\in\mathcal{D}$ and the operators $A_z$ and $A_{\oz}$ depend on $\alpha$ and $n$). However, there is no particular values for $n$ and $\alpha$ for which the commutator of $A_z$ and $A_{\oz}$ becomes the exact $su(1,1)$ commutation relation. This is due to the fact that the basis $B$ in (\ref{U3}) or $B_{\eta}$ in (\ref{U9}) cannot be obtained from the basis $B_n^{\alpha}$ of (\ref{B1}) by adjusting the parameters $n$ and $\alpha$. However, between the spaces $\mathcal{FB}_{\eta}$ of Section 2 and $\Lambda_n^{\alpha}(\mathcal{D})$, we can draw an isometry as follows.
\end{remark}
\subsection{An isometry between $\mathcal{FB}_{\eta}$ and $\Lambda_n^{\alpha}(\mathcal{D})$}
Since $B_{\eta}$ in (\ref{U9})
is an orthonormal basis of $\mathcal{FB}_{\eta}$, in the CS (6.1) we replace the orthonormal basis $\{|e_m\rangle~|~m\in\mathbb{N}\}$ by $\{|p_m\rangle~|~m\in\mathbb{N}\}$ of $\mathcal{FB}_{\eta}$. 
The set $B_n^{\alpha}$ of (\ref{B1}) is an orthonormal basis of $\Lambda_{n}^{\alpha}(\mathcal{D}).$
Write the CS as
$$|z,n,\alpha\rangle=\mathcal{N}(z,\oz)^{-\frac{1}{2}}\sum_{m=0}^{\infty}
\frac{P_{m,n}^{\alpha}(z,\oz)}{\sqrt{A_{\alpha}(m,n)}}|p_m\rangle\in FB_{\eta}.$$
Now, through the resolution of the identity $F_n^{\alpha}(z)=\langle\xi|z,n,\alpha\rangle$, for all $|\xi\rangle\in \mathcal{FB}_{\eta}$ are square integrable with respect to the resolution of the identity measure, and thereby, defines elements in  $\Lambda_{n}^{\alpha}(\mathcal{D}).$ Again through the resolution of the identity
$$W:\mathcal{FB}_{\eta}\longrightarrow \Lambda_{n}^{\alpha}(\mathcal{D})\quad |\xi\rangle\mapsto F_n^{\alpha}\quad by \quad W(|\xi\rangle)(z)= \langle\xi|z,n,\alpha\rangle$$ is a linear isometry, in particular $W(p_m)=B_{n,m}^{\alpha}$,
where
\begin{equation}\label{X1}
\langle\xi|z,n,\alpha\rangle=\mathcal{N}(z,\oz)^{-\frac{1}{2}}\sum_{m=0}^{\infty}
\frac{P_{m,n}^{\alpha}(z,\oz)}{\sqrt{A_{\alpha}(m,n)}}\sqrt{\frac{(2\eta)_n}{n!}}\xi^n,
\end{equation}
which is an absolutely convergent series (see lemma \ref{T3}). A note is in order about the closed form of (\ref{X1}). Due to the square root in the term $\sqrt{A_{\alpha}(m,n)}$, we faced difficulty in using the known formulas to get a closed form for (\ref{X1}). However, for a very specific choice of $n$ and $\eta$ we were able to find a closed form (see again lemma \ref{T2} and the appendix).
\subsection{Overlap of the CS, lower symbols and Berezin transform}
Using lemma (\ref{T1}) we can compute the overlap of two CS as
\begin{eqnarray}\label{E227}
\langle z,\alpha,n|w,\alpha,n\rangle&=&[{\N_n(z,\oz)\N_n(w,\overline{w})}]^{-1/2}
\sum_{m=0}^{\infty}\frac{\overline{\PP(z,\oz)}\PP(w,\overline{w})}{A_{\alpha}(m,n)}\\
&=&\frac{\pi\left[(1-z\oz)(1-w\overline{w})\right]^{({\alpha+2})/{2}}}{(2n+\alpha+1)}E_n^{\alpha}(\oz,w).\notag
\end{eqnarray}
Now, from (\ref{E8}) and (\ref{E227}), the lower symbol of $A_f$ can be computed as
\begin{eqnarray*}
\check{f}=\langle z,n,\alpha|A_f|z,n,\alpha\rangle&=&\int_{\D}\N_n(w,\overline{w})f(w)|\langle z,n,\alpha|w,n,\alpha\rangle|^2 d\mu(w)\\
&=&\frac{\pi\,(1-z\oz)^{\alpha+2}}{(2n+\alpha+1)}\int_{\D}
\left[E_n^{\alpha}(\oz,w)\right]^2f(w)\,d\mu(w).
\end{eqnarray*}
Thereby the Berezin transform can be written as
\begin{equation}\label{BB1}
B_n^{\alpha}[f](z)=\langle z,n,\alpha|A_f|z,n,\alpha\rangle=\frac{\pi(1-z\oz)^{\alpha+2}}{(2n+\alpha+1)}\int_{\D}
\left[E_n^{\alpha}(\oz,w)\right]^2f(w)\,d\mu(w).
\end{equation}
Equation (\ref{BB1}) can be considered as a more generalized Berezin transform for the unit disc in a sense that $n=0$, that is
$$E_0^{\alpha}(\oz,w)=\frac{(\alpha+1)}{\pi(1-w\oz)^{\alpha+2}},$$
leads to the Berezin transform of the weighted Bergman space, $\mathcal{A}^2_{\alpha}(\D)$, \cite{SD}
\begin{equation}\label{BB2}
B_0^{\alpha}[f](z)=\langle z,0,\alpha|A_f|z,0,\alpha\rangle=\frac{(\alpha+1)}{\pi}\int_{\D}\frac{(1-z\oz)^{\alpha+2}}
{(1-w\oz)^{4+2\alpha}}f(w)d\mu(w)
\end{equation}
and the values $n=0$ and $\alpha=0$, leads to the standard Berezin transform of the unit disc (see \cite{En}, pp-68), namely
\begin{equation}
B[f](z)=\frac{1}{\pi}(1-|z|^2)^2\int_{\D}\frac{f(w)}{|1-\oz w|^4}d\mu(w).
\end{equation}
\section{conclusion}
\noindent Along the lines of the general scheme of CS quantization, using  CS, built with 2D-Zernike polynomials, we have quantized the complex unit disc. In assisting the computations of lower symbols, upper symbols, and the Berezin transform we have provided some interesting summation and integral formulas for 2D-Zernike polynomials. The unit disc is a natural phase space for quantum systems with $su(1,1)$ symmetry, and thereby, while quantizing the unit disc it is natural to expect similar  symmetry. However, as a remarkable feature we have shown that, for $z\in\mathcal{D}$, the operators $A_z$ and $A_{\oz}$ arising from the 2D-Zernike polynomial quantization do not obey such symmetry.
\vskip0.1true in
\noindent Using the CS, constructed in the present work, one may consider studying modular structures on $\D$ as it was done with the complex Hermite polynomials \cite{AFG}. Further, similar to \cite{In}, one can study the spectral properties of Cauchy transform on $L_{\alpha}^2(\D, \frac{i}{2}dz\wedge d\oz)$ because its comparable analysis to the one developed in section 5. For $\PP(z,\oz)$, by setting $n\geq m$, namely $n=m+s$, we can obtain a family of reproducing kernel Hilbert subspaces of  $L_{\alpha}^2(\D, \frac{i}{2}(1-z\oz)^{\alpha}dz\wedge d\oz)$ similar to the earlier work of \cite{ABG} for the complex plane $\mathbb{C}$ and thus a similar study to the work of \cite{ABG} can be focused on the disc $\D$.  Some of these issues will be tackle in a future work.
\section{Acknowledgements}
\medskip
\noindent Partial financial support of this work under Grant No. GP249507 from the
Natural Sciences and Engineering Research Council of Canada
 is gratefully acknowledged by one of us (NS). The authors like to thank the anonymous referees for their valuable insights  improving the present work. 
\section{Appendix}
\subsection{Proof of Lemma \ref{T1}}
\noindent For the case $n=0$ we have, using \eqref{E9}, 
\begin{align}\label{app1}
\displaystyle{\sum_{m=0}^{\infty}\frac{P_{m,0}^{\alpha}(z,\oz)\overline{P_{m,0}^{\alpha}(w,\overline{w})}}
{A_{\alpha}(m,0)}=\dfrac{(\alpha+1)}{\pi}
\sum_{m=0}^\infty \dfrac{(\alpha+2)_m}{m!} (\overline{w} z)^m
=
\frac{(\alpha+1)}
{\pi(1-\overline{w}z)^{\alpha+2}}}.
\end{align}
by means of the Pochhammer identity
$$(1+\alpha+m)(1+\alpha)_m=(1+\alpha) (2+\alpha)_m,$$
and the summation
\begin{align}\label{app2}
\sum_{m=0}^\infty \dfrac{(\alpha+2)_m}{m!} (\overline{w} z)^m=(1-\overline{w}z)^{-2-\alpha}.
\end{align}
In general,
\begin{align}\label{app3}
\sum_{m=0}^{\infty}\frac{\PP(z,\oz)\overline{\PP(w,\overline{w})}}{A_{\alpha}(m,n)}&=\dfrac{(\alpha+1)_n}{\pi\, n!}(\oz \, w)^n\sum_{m=0}^\infty \dfrac{(m+n+\alpha+1)\, (\alpha+1)_m}{m!}\, (z\cdot \overline{w})^m\notag\\
&\times {}_2F_1(-m,-n;\alpha+1; 1- \frac{1}{z\overline{z}})\, {}_2F_1(-m,-n;\alpha+1; 1- \frac{1}{w\overline{w}})\notag\\
&=n\dfrac{(\alpha+1)_n}{\pi\, n!}(\oz \, w)^n\sum_{m=0}^\infty \dfrac{(\alpha+1)_m}{m!}\, (z\, \overline{w})^m\notag\\
&\times {}_2F_1(-m,-n;\alpha+1; 1- \frac{1}{z\overline{z}})\, {}_2F_1(-m,-n;\alpha+1; 1- \frac{1}{w\overline{w}})\notag\\
&+\dfrac{(\alpha+1)\,(\alpha+1)_n}{\pi\,n!}(\oz \, w)^n\sum_{m=0}^\infty \dfrac{(\alpha+2)_m}{\Gamma(m+1)}\, (z\, \overline{w})^m\notag\\
&\times {}_2F_1(-m,-n;\alpha+1; 1- \frac{1}{z\overline{z}})\, {}_2F_1(-m,-n;\alpha+1; 1- \frac{1}{w\overline{w}})\notag\\
&=\dfrac{n\,(\alpha+1)_n}{\pi\,n!}(\oz \, w)^n\,S_1+\dfrac{(\alpha+1)\,(\alpha+1)_n}{\pi\, n!}(\oz\, w)^n\, S_2.
\end{align}
For the first sum $S_1$, we use the following identity (note we corrected here the typo of the formula as given in Reference \cite{cap}, formula  (3)), 
\begin{align}\label{app4}
\sum_{m=0}^{\infty}\dfrac{(1+\alpha)_m\,z^m}{m!}&\,{}_2F_1(-m,\beta;1+\alpha;\nu)\, {}_2F_1(-m,b;1+\alpha;u)\notag\\
&=\dfrac{(1-z)^{\beta+b-\alpha-1}}{(1-z+\nu z)^{\beta}(1-z+u z)^{b}} {}_2F_1\left(\beta,b;1+\alpha;\dfrac{z\nu u}{(1-z+\nu z)(1-z+u z)}\right)
\end{align}
where $|u|<1,\,|\nu|<1$ and $|z|<1$, thus
\begin{align}\label{app5}
S_1&=\dfrac{\left(\oz-\overline{w}\right)^{n} \left(w-z\right)^{n}}{\oz^n\,w^n\,(1-z \overline{w})^{2n+1+\alpha}} {}_2F_1\left(-n,-n;1+\alpha;\dfrac{\left(z\oz -1\right) \left(w\overline{w}-1\right)}{\left(\oz-\overline{w}\right) \left(w-z\right)}\right)
\end{align}
Using the terminated series representation of the Gauss hypergeometric function ${}_2F_1(-n,\beta;\gamma;z)$ and collecting the terms, we obtain
\begin{align}\label{app6}
\lim_{(w,\overline{w})\rightarrow (z,\oz)}&\left[\dfrac{n\,(\alpha+1)_n}{\pi\,n!}\, (\oz \, w)^n\times S_1\right]=\dfrac{n}{\pi}(1-z\oz)^{-1-\alpha}.
\end{align}
For the second sum $S_2$, we use the following identity (\cite{sri1}, formula 2.6)
\begin{align}\label{app7}
&\sum_{m=0}^\infty \dfrac{(\lambda)_m}{m!}{}_2F_1(-m,\mu;\alpha+1;x)\,{}_2F_1(-m,\nu;\beta+1;y)z^m \notag\\
&=\dfrac{(1-z)^{\mu-\lambda}}{(1-z+xz)^{\mu}}\sum_{m=0}^\infty \dfrac{(\lambda)_m\, (\nu)_m}{m!\, (\beta+1)_m}\left(\dfrac{yz}{z-1}\right)^mF_1\left(\mu,-m,\alpha-\lambda+1;\alpha+1;\dfrac{x}{1-z+xz},\dfrac{xz}{1-z+xz}\right)\notag\\
\end{align}
where $F_1$ is the first Appell series of two variables \cite{sri2}. Thus, we have
\begin{align}\label{app8}
S_2&=(1-z\overline{w})^{-n-\alpha-2}\left(1-\dfrac{\overline{w}}{\oz}\right)^n\, \sum_{k=0}^n\dfrac{(-n)_k (\alpha+2)_k}{k! (\alpha+1)_k}\left(\dfrac{z(w\overline{w}-1)}{w(z\overline{w}-1}\right)^k\notag\\
&\times F_1\left(-n,-k,-1;1+\alpha;\dfrac{z\oz-1}{z(\oz-\overline{w})},\dfrac{\overline{w}(z\oz-1\overline{w}}{\oz-\overline{w}}\right)
\end{align}
where we used the fact that $(-n)_k=0$ for $k>n$. Now, by means of the double series representation of the first Appell function, we have
\begin{align}\label{app9}
\lim_{(w,\overline{w})\rightarrow (z,\oz)}&\left[\dfrac{(\alpha+1)(\alpha+1)_n}{\pi\, n!}(\oz\, w)^n\times S_2\right]=\dfrac{(n+1+\alpha+nz\oz)}{\pi}\,(1-z\oz)^{-2-\alpha}.
\end{align}
Consequently, from \eqref{app6} and \eqref{app9}, we have for \eqref{app3} in the case of $z=w$ that
 \begin{align}\label{app10}
\sum_{m=0}^{\infty}\frac{\PP(z,\oz)\overline{\PP(z,\oz)}}{A_{\alpha}(m,n)}&=\dfrac{1}{\pi}\dfrac{(2n+1+\alpha)}{(1-z\, \oz)^{\alpha+2}}.
\end{align}
In general, for $z\neq w$, we have using
\begin{align}\label{app11}
F_1\bigg(-n,-k,-1;&1+\alpha;\dfrac{z\oz-1}{z(\oz-\overline{w})},\dfrac{\overline{w}(z\oz-1\overline{w}}{\oz-\overline{w}}\bigg)\notag\\
&=\dfrac{\left((\alpha+n-1)\overline{w}-(1+\alpha+n\,z\,\overline{w})\oz\right)}{(1+\alpha)(\overline{w}-\oz)}{}_2F_1\left(-n,-k;2+\alpha;\dfrac{z\oz-1}{z(\oz-\overline{w})}\right)\notag\\
&-\dfrac{k\,n\oz\,(z\oz-1)\,(z\overline{w}-1)}{(1+\alpha)(2+\alpha)z(\overline{w}-\oz)^2}\,{}_2F_1\left(1-n,1-k;3+\alpha;\dfrac{z\oz-1}{z(\oz-\overline{w})}\right),
\end{align}
and the identity
\begin{align}\label{app12}{}_2F_1(1+a,1+b;1+c;z)=\dfrac{c}{bz}\left[{}_2F_1(1+a,b;c;z)-{}_2F_1(a,b;c,z)\right],
\end{align}
that
\begin{align}\label{app122}
F_1\bigg(-n,-k,-1;1+\alpha;\dfrac{z\oz-1}{z(\oz-\overline{w})},\dfrac{\overline{w}(z\oz-1\overline{w}}{\oz-\overline{w}}\bigg)&=\dfrac{n\,\oz\,(z\,\overline{w}-1)}{(1+\alpha)\,(\oz-\overline{w})}{}_2F_1\left(1-n,-k;2+\alpha;\dfrac{z\oz-1}{z(\oz-\overline{w})}\right)\notag\\
&+\dfrac{(1+\alpha+n)}{1+\alpha}\,{}_2F_1\left(-n,-k;2+\alpha;\dfrac{z\oz-1}{z(\oz-\overline{w})}\right).
\end{align}
Consequently,
\begin{align}\label{app13}
S_2&=\dfrac{(1+\alpha+n)(\oz-\overline{w})^n}{(1+\alpha)\oz^n(1-z\overline{w})^{n+\alpha+2}}\sum_{k=0}^n\dfrac{(2+\alpha)_k\,(-n)_k}{k!\,(1+\alpha)_k}\,\left(\dfrac{z(w\overline{w}-1)}{w(z\overline{w}-1}\right)^k{}_2F_1\left(-n,-k;2+\alpha;\dfrac{z\oz-1}{z(\oz-\overline{w})}\right)\notag\\
&-\dfrac{n(\oz-\overline{w})^{n-1}}{(1+\alpha)\oz^{n-1}(1-z\overline{w})^{n+\alpha+1}}\sum_{k=0}^n\dfrac{(2+\alpha)_k\,(-n)_k}{k!\,(1+\alpha)_k}\,\left(\dfrac{z(w\overline{w}-1)}{w(z\overline{w}-1}\right)^k{}_2F_1\left(1-n,-k;2+\alpha;\dfrac{z\oz-1}{z(\oz-\overline{w})}\right)\notag\\
\end{align}
Finally, by means of the identity
\begin{align}\label{app14}
\sum_{k=0}^\infty\dfrac{(a)_k\,(b')_k}{k!\,(c')_k}\,t^k\,{}_2F_1\left(-k,b;a;\dfrac{x}{x-1}\right)&=(1-x)^{b}F_2(a;b,b';a,c';x,t),
\end{align}
the proof of the lemma follows.

\subsection{Proof of Lemma \ref{T2}} For $n=0$,
\begin{align*}
\sum_{m=0}^\infty\dfrac{P_{m,n}^\alpha (z,\overline{z})}{\sqrt{A_\alpha(m,n)}}\sqrt{\dfrac{(2\eta)_m}{m!}}\xi^m
&=\dfrac{1}{\sqrt{\pi}}
\sum_{m=0}^\infty \dfrac{\sqrt{{(m+\alpha+1)\,(1+\alpha)_m(2\eta)_m}}}{m!}{(\xi\,z)^m}\notag\\
&=\sqrt{\dfrac{a+1}{\pi}}
\sum_{m=0}^\infty \dfrac{(2+\alpha)_m}{m!}{(\xi\,z)^m}=\sqrt{\dfrac{a+1}{\pi}}\dfrac{1}{(1-\xi\,z)^{\alpha+2}},
\end{align*}
thus if $\eta=({2+\alpha})/{2}$
\begin{align*}
\sum_{m=0}^\infty\dfrac{P_{m,n}^\alpha (z,\overline{z})}{\sqrt{A_\alpha(m,n)}}\sqrt{\dfrac{(2\eta)_m}{m!}}\xi^m
&=\sqrt{\dfrac{a+1}{\pi}}(1-\xi\,z)^{-\alpha-2}.
\end{align*}
In general,
\begin{align*}
\sum_{m=0}^\infty\dfrac{P_{m,n}^\alpha (z,\overline{z})}{\sqrt{A_\alpha(m,n)}}\sqrt{\dfrac{(2\eta)_m}{m!}}\xi^m
&=\sqrt{
\dfrac{(1+\alpha)_n}{\pi\, \,n!}}\sum_{m=0}^\infty\sqrt{
{(2\eta)_m\,(1+\alpha)_m}}\,\dfrac{\xi^m}{m!}\sqrt{m+n+\alpha+1}\,P_{m,n}^\alpha (z,\overline{z}),
\end{align*}
using the definition of $P_{m,n}^\alpha (z,\overline{z})$, equation (2.7) in \cite{Wu3},
\begin{align*}
\sum_{m=0}^\infty\dfrac{P_{m,n}^\alpha (z,\overline{z})}{\sqrt{A_\alpha(m,n)}}\sqrt{\dfrac{(2\eta)_m}{m!}}\xi^m
&=\sqrt{
\dfrac{(1+\alpha)_n}{\pi\, \,n!}}\dfrac{(-1)^n}{(\alpha+1)_n(1-z\overline{z})^\alpha}\notag\\
&\times\dfrac{\partial^n}{\partial z^n}(1-z\overline{z})^{n+\alpha}
\sum_{m=0}^\infty\sqrt{
{(2\eta)_m\,(1+\alpha)_m}}\,\dfrac{(z\xi)^m}{m!}\sqrt{m+n+\alpha+1}.
\end{align*}
However,
$$\sqrt{m+n+\alpha+1}<\sqrt{m+\alpha+1}+\sqrt{n},\quad for\quad \alpha>-1$$
Thus, for $\eta>(\alpha+2)/2$, $\alpha>-1$,
\begin{align*}
\sum_{m=0}^\infty\sqrt{
{(2\eta)_m\,(1+\alpha)_m}}&\,\dfrac{(z\xi)^m}{m!}\sqrt{m+n+\alpha+1}
\notag\\
&<({\alpha+1})\sum_{m=0}^\infty\sqrt{
{(2\eta)_m\,(2+\alpha)_m}}\,\dfrac{(z\xi)^m}{m!}+\sqrt{n}\sum_{m=0}^\infty\sqrt{
{(2\eta)_m\,(1+\alpha)_m}}\,\dfrac{(z\xi)^m}{m!}.
\end{align*}
and consequently
\begin{align*}
\sum_{m=0}^\infty\sqrt{
{(2\eta)_m\,(1+\alpha)_m}}&\,\dfrac{(z\xi)^m}{m!}\sqrt{m+n+\alpha+1}
<\dfrac{1}{\alpha+1}\sum_{m=0}^\infty
(2\eta)_m\,\dfrac{(z\xi)^m}{m!}+\sqrt{n}\sum_{m=0}^\infty
(2\eta)_m\,\dfrac{(z\xi)^m}{m!}\notag\\
&=\left({\alpha+1}+\sqrt{n}\right)\sum_{m=0}^\infty
\,\dfrac{(2\eta)_m(z\xi)^m}{m!}=\left({\alpha+1}+\sqrt{n}\right)\left(1-z\xi\right)^{-2\eta}
\end{align*}
Finally,
\begin{align*}
\sum_{m=0}^\infty\bigg|\dfrac{P_{m,n}^\alpha (z,\overline{z})}{\sqrt{A_\alpha(m,n)}}\sqrt{\dfrac{(2\eta)_m}{m!}}\xi^m\bigg|
&=\sqrt{
\dfrac{(1+\alpha)_n}{\pi\, \,n!}}\dfrac{1}{(\alpha+1)_n(1-z\overline{z})^\alpha}\notag\\
&\times\dfrac{\partial^n}{\partial z^n}(1-z\overline{z})^{n+\alpha}
\sum_{m=0}^\infty\sqrt{
{(2\eta)_m\,(1+\alpha)_m}}\,\dfrac{(z\xi)^m}{m!}\sqrt{m+n+\alpha+1}\notag\\
&<\sqrt{
\dfrac{(1+\alpha)_n}{\pi\, \,n!}}\dfrac{\left({\alpha+1}+\sqrt{n}\right)}{(\alpha+1)_n(1-z\overline{z})^\alpha}\dfrac{\partial^n}{\partial z^n}\left(\dfrac{(1-z\overline{z})^{n+\alpha}}{
(1-z\xi)^{2\eta}}\right).
\end{align*}
where $\eta\geq(\alpha+2)/2.$
Thus, the series
\begin{align*}
\sum_{m=0}^\infty\dfrac{P_{m,n}^\alpha (z,\overline{z})}{\sqrt{A_\alpha(m,n)}}\sqrt{\dfrac{(2\eta)_m}{m!}}\xi^m,
\end{align*}
absolutely convergent.

\subsection{Proof of Lemma \ref{T3}} Note first
\begin{align*}
\int_{z\oz\leq 1}&f(\theta) \,\overline{P_{l,n}^{\alpha}\,(z,\oz)}\PP(z,\oz)d\mu_\alpha(z,\oz) 
=\int_0^{2\pi}f(\theta)  e^{i(m-l)\theta}d\theta\notag\\
&\times\int_0^1 r(1-r^2)^\alpha r^{l+2n+m} {}_2F_1(-l,-n;\alpha+1; 1- \frac{1}{r^2}) {}_2F_1(-m,-n;\alpha+1; 1- \frac{1}{r^2})dr.
\end{align*}
However
\begin{align*}
\int_0^1 r(1-r^2)^\alpha r^{l+2n+m}& {}_2F_1(-l,-n;\alpha+1; 1- \frac{1}{r^2}) {}_2F_1(-m,-n;\alpha+1; 1- \frac{1}{r^2})dr\notag\\
&=\dfrac{m!\,n!}{2(m+n+\alpha+1)(1+\alpha)_m(1+\alpha)_n}\delta_{l,m}.
\end{align*}
and 
$$\int_0^{2\pi}f(\theta)  e^{i(m-l)\theta}d\theta=\int_0^{2\pi}f(\theta) d\theta=C~\mbox{(independent of $m$ and $l$ where $m=l$)},$$
Then 
\begin{align*}
\int_{z\oz\leq 1}f(\theta) \,\overline{P_{l,n}^{\alpha}\,(z,\oz)}\PP(z,\oz)d\mu_\alpha(z,\oz)= \dfrac{C m!\,n!}{2(m+n+\alpha+1)(1+\alpha)_m(1+\alpha)_n}\delta_{l,m},
\end{align*}
Furthermore,
\begin{align*}
A_{f(\theta)}=\sum_{m=0}^\infty &\sum_{l=0}^\infty\dfrac{|e_m\rangle\langle e_l|}{\sqrt{A_\alpha(m,n)A_\alpha(l,n)}}\int_{z\oz\leq 1}f(\theta) \,\overline{P_{l,n}^{\alpha}\,(z,\oz)}\PP(z,\oz)d\mu_\alpha(z,\oz)\notag\\
&=\frac{C}{2}\sum_{m=0}^\infty \sum_{l=0}^\infty\dfrac{|e_m\rangle\langle e_l|}{\sqrt{A_\alpha(m,n)A_\alpha(l,n)}}\dfrac{m!\,n!}{(m+n+\alpha+1)(1+\alpha)_m(1+\alpha)_n}\delta_{l,m}.\notag\\
&=\frac{C}{2\pi}\sum_{m=0}^\infty |e_m\rangle\langle e_m|=\frac{C}{2\pi} I_{\mathfrak{H}},\qquad \mbox{where $C=\int_0^{2\pi}f(\theta) d\theta$.}
\end{align*}


\begin{thebibliography}{XXXX}
\bibitem{En}{\em Encyclopaedia of Mathematics. Vol. 3. D-Feynman measure}, Eds. Hazewinkel, M., Kluwer Academic publications, Netherlands, 2001.

\bibitem{Ali} Ali, S.T., Antoine, J-P., Gazeau, J-P.,
{\em Coherent States, Wavelets and Their Generalizations},
Springer, New York, 2000.

\bibitem{ABG}Ali, S.T., Bagarello, F., Gazeau,J-P., {\em Quantization from reproducing kernel spaces}, Ann. Phys. {\bf 332} (2012), 127-142.

\bibitem{AFG} Ali, S.T., Bagarello, F., Honnouvo, G., {\em Modular structures on trace class operators and applications to Landau levels}, J.Phys.A: Math. Theor.,{\bf 43}, (2010), 105202.

\bibitem{AE} Ali, S.T., Engli\v{s}, M., {\em Quantization methods: a guide for physicists and analysts}, Rev. Math. Phys. {\bf 17} (2005), 391-490.

\bibitem{AGH}Aremua, I., Gazeau, J-P., Hounkonnou, M.N., {\em Action-angle coherent states for quantum systems with cylindrical phase space}, J. Phys. A. {\bf 45} (2012), 335302 (16pp).

\bibitem{As} Askour, N., Mouayn, Z., {\em Probality distributions attached to generalized Bergman spaces on the Poincar\'e disc}, arXiv:1003.4323.

\bibitem{Ba}Bailey,  W. N. , \emph{Generalized hypergeometric series}, Cambridge University Press, Cambridge, 1935.

\bibitem{El}Elwassouli, F., Ghanmi, A., Intissar, A., Mouayn, Z., {\em Generalized second Bargmann transforms associated with the hyperbolic Landau levels in the Poincar\'e disc}, Ann. Henri Poincar\'e. {\bf 13} (2012), 513-524.

\bibitem{Nic} Cotfas, N., Gazeau, J-P., Gorska, K., {\em Complex and real Hermite polynomials and related quantizations}, J. Phys.A: Math. Theor. {\bf 43} (2010), 305304.

\bibitem{cap} Capelas de Oliveira, E., \emph{On generating functions} Il Nuovo Climento {\bf 107 B}, (1992) 59-64.

\bibitem{ES} Egorow, Y.V., Shubin, M.A., {\em Linear partial differential equations. Foundations of the classical theory}, (2nd Edition), Springer-Verlag, Berlin, 1998.

\bibitem{sri2} Erd\'elyi, A. et al., \emph{Higher transcendental functions}, Vol. II, McGraw-Hill, New York,
1953. 

\bibitem{E} Engli\v{s}, M., {\em Analytic continuation of weighted Bergman kernels}, J. Math. Pures Appl (9). {\bf 94} (2010), 622-650.

\bibitem{Gaz} Gazeau, J-P., {\em Coherent states in quantum physics}, Wiley-VCH, Berlin (2009).

\bibitem{Ga} Gazeau, J-P., Szafraniec, F.H., {\em Holomorphic Hermite polynomials and non-commutative plane}, J. Phys. A: Math. Theor. {\bf 44} (2011), 495201.

\bibitem{Hall} Hall, B.C., \emph{Holomorphic methods in analysis and mathematical physics}. First Summer School in
Analysis and Mathematical Physics (Cuernavaca, Morelos, 1998), Contemp. Math. {\bf 260} 1, Am. Math.
Soc. Providence RI, 2000.

\bibitem{In} Intissar, A., Intissar A., {\em Spectral properties of the Cauchy transform on $L_2 (\mathbb C, e^{-\vert z \vert^2} \lambda (z))$}, J. Math. Anal. and Applications {\bf 313} (2006), 400-418.

\bibitem{Ko} Kolaski, C.J., {\em Isometries of weighted Bergman spaces}, Can. J. Math. {\bf 34} (1982), 910-915.

\bibitem{Mo} Mouayn, Z., {\em Coherent states quantization for generalized Bargmann spaces with formulae for their attached Berezin transform in terms of the Laplacian on $\mathbb{C}^n$}, J. Fourier. Anal. Appl. {\bf 18}, (2012), 609-625.

\bibitem{Per}Perelomov, A., {\em Generalized coherent states and their applications}, Springer-Verlag, Berlin (1986).

\bibitem{sri1}Srivastava, H. M.,  \emph{An extension of the Hille-Hardy formula,} Math. Comput. {\bf 23} (1969) 305-311. 

\bibitem{SD}Stroethoff, K., Zheng, D., {\em Bounded Toeplitz products on weighted Bergman spaces}, J. Operator theory, {\bf 59} (2008) 277-308.

\bibitem{Tor} Torre, A., {\em Generalized Zernike or disc polynomials: An application in quantum optics}, J. Comput. Appl. Math. {\bf 222} (2008) 622-644.

\bibitem{Wu3} W\"unsche, A., {\em Generalized Zernike or disc polynomials}, J. Comput. Appl. Math. {\bf 174} (2005) 135-163.


\end{thebibliography}
\end{document}